# Deep Koopman Operator-informed Safety Command Governor for Autonomous Vehicles

Hao Chen, Xiangkun He, Shuo Cheng, Chen Lv, *Senior Member, IEEE*

***Abstract*—Modelling of nonlinear behaviors with physical-based models poses challenges. However, Koopman operator maps the original nonlinear system into an infinite-dimensional linear space to achieve global linearization of the nonlinear system through input and output data, which derives an absolute equivalent linear representation of the original state space. Due to the impossibility of implementing the infinite-dimensional Koopman operator, finite-dimensional kernel functions are selected as an approximation. Given its flexible structure and high accuracy, deep learning is initially employed to extract kernel functions from data and acquire a linear evolution dynamic of the autonomous vehicle in the lifted space. Additionally, the control barrier function (CBF) converts the state constraints to the constraints on the input to render safety property. Then, in terms of the lateral stability of the in-wheel motor driven vehicle (IMDV), the CBF conditions are incorporated with the learned deep Koopman model. Because of the linear fashion of the deep Koopman model, the quadratic programming (QP) problem is formulated to generate the applied driving torque with minimal perturbation to the original driving torque as a safety command governor. In the end, to validate the fidelity of the deep Koopman model compared to other mainstream approaches and demonstrate the lateral improvement achieved by the proposed safety command governor, data collection and safety testing scenarios are conducted on a hardware-in-the-loop (HiL) platform.**

***Index Terms*—Nonlinear system modelling, deep Koopman operator, control barrier function, safety command governor, autonomous vehicles.**

## I. Introduction

NONLINEAR behaviors are difficult to characterize with physical-based models, and even if it is possible, the expressions are quite complicated [1]–[3]. Besides, many well-established control techniques are based on linear systems or input affine systems, whereas mechatronic systems, particularly those classes with high nonlinear dynamics, such as vehicles, cannot satisfy such prerequisites or assumptions. Thereby, this limits the application and extension of methods. Data-driven modelling methodology provides a tutorial on these issues. The data-driven method does not rely on the internal structure and parameters of the control plant but rather on the input and output data, which can effectively characterize the nonlinear components and improve the model fidelity of the system [4]–[7].

Hao Chen, Xiangkun He, and Chen Lv are with the School of Mechanical and Aerospace Engineering, Nanyang Technological University, 639798, Singapore, e-mails: chen.h@ntu.edu.sg, xiangkun.he@ntu.edu.sg, lyuchen@ntu.edu.sg. Shuo Cheng is with the Institute of Industrial Science, the University of Tokyo, Tokyo, 153-0041, Japan. e-mail: cshuo@iis.u-tokyo.ac.jp

Corresponding author: C. Lv

Koopman operator maps the original nonlinear system into an infinite-dimensional linear space to achieve global linearization of the nonlinear system through experimental or simulation data [8]–[11]. Unlike the local linearization method, Koopman operator theory does not constrain the system behavior close to the working point or expect an accurate physical model. In contrast, it achieves linear representation by compressing the nonlinear features in the lifted space, and the Koopman representation has marginally better performance than other nonlinear modelling methods, such as SINDy and NARX [12]. However, the infinite-dimensional Koopman operator cannot be implemented; instead, finite-dimensional basis functions are chosen to approximate the Koopman operator, that is, the proper choice of basis functions used for the linear lifting. Therefore, the promise of the Koopman approach is to take the tools developed for linear systems and apply them to the dynamical system defined by the Koopman operator, thus obtaining a linear approximation of a nonlinear system without directly linearizing around a particular fixed point.

Here are two mainstreams for linear lifting: dynamic mode decomposition (DMD) and extended dynamic mode decomposition (EDMD). DMD directly employs linear functions as kernel functions, which are still linear transformations and cannot characterize nonlinear systems [13]. EDMD, employs nonlinear functions such as radial basis functions and polynomials as kernel functions to achieve the mapping from low-dimensional nonlinear space to high-dimensional linear space, hence offering a novel approach for the prediction and control of nonlinear systems. Cibulka et al. chose polynomial basis functions, consisting of monomials of the state vector elements, to approximate the Koopman operator. In terms of the single-track vehicle dynamics, the EDMD approach shows good accuracy, with RMSE equal to 11.1% [14]. However, EDMD's kernel function requires prior knowledge and numerous trials, which are also dependent on the specific control object and lack of generalization [15], [16].

Deep learning is a machine learning technique that teaches computers to do what comes naturally to humans. Models are trained by using a large set of labeled data and neural network architectures that contain many layers and then achieve state-of-the-art accuracy, sometimes exceeding human-level performance [17], [18]. Through the training and learning of deep neural networks, kernel functions can be directly extracted from data to generate Koopman operators [19]. Rongyao Wang et al. developed a deep neural network as the Koopman operator to identify the augmented dynamics of scaled vehicles. The explainable and linearized system representation in a high-dimensional latent space is proven to be flexible and robust in different simulation scenarios [20]. Haojie Shi proposed an end-to-end deep learning framework to learn the Koopman operator and evaluated the approach in several nonlinear dynamic systems. It reduces prediction error by an order of magnitude compared to existing Koopman-based methods and deep learning-based approaches [21].

Safety is a paramount concern in the realm of mechatronics system design, necessitating the prevention of constraint violations. Concretely, the safety property guarantees the forward invariance that any trajectory starts inside a set satisfying state, and input constraints will never reach the complement of the set [22]. The states and inputs are defined as decisive variables for the optimization problem, and the optimal solution is found in the admissible set, such as model predictive control (MPC), but the computational effort is often expensive, especially for nonlinear systems [23], [24]. The control barrier function (CBF) maps the state constraints onto the constraints on the input [25]–[27]. The feasible control input does not violate the original constraint and ensures the forward invariance of the associated set, i.e., safety property. This enables the CBF method to be suitable for complex objectives and constraints. Meanwhile, if the system is confined to linear or nonlinear affine control form, the CBF conditions on safety are affine in the control input. Thus, we can consider a quadratic programming (QP) optimization problem to develop the controller with excellent solution performance [22], [28]. The deep Koopman operator has identified a linear system in the lifted space. Motivated by this, in this paper, we employ the deep neural network to obtain a linear evolution dynamic of the autonomous vehicle in the lifted space. The CBF conditions are incorporated with the learned Koopman model as the constraints on the control input. The formulation of the QP problem is to generate applied driving torque with minimal perturbation to the original driving torque. This results in a command governor rendering safety property. In addition, various scenarios are set on the HiL platform of dSPACE/SCALEXIO and Logitech G29 to facilitate comprehensive comparisons and rigorous verification of both the precise model fidelity and the lateral stability enhancement offered by the proposed safety command governor.

The main contributions of this study are concluded as:

1) A novel framework of deep Koopman operator-informed safety governor is introduced that incorporates data-driven modelling methods into safety-critical control theory. Due to the high model fidelity and the linearity in the lifted space via the Koopman operator, a QP problem is formulated as the command governor to improve the safety of autonomous vehicles.
2) The control barrier functions are defined as the boundaries of lateral stability regions, which ensure the forward invariant of vehicle states by input constraints, that is, maintaining lateral motions in the safe set.
3) The experimental tests are conducted on the hardware-in-the-loop platform- dSPACE/SCALEXIO, to verify the efficacy of the proposed safety command governor in terms of vehicle lateral stability.

The remaining paper is organized as follows: The information about the deep Koopman operator and control barrier function is introduced in Section II. The framework of deep Koopman operator-informed safety command governor, including the neural network as the embedding functions, Koopman operator that lifts the original state space into a high-dimensional linear space, and the CBF conditions in terms of lateral stability, is discussed in Section III. Section IV conducts the validation of the model fidelity of the deep Koopman operator method on the HiL platform. Besides, the lateral stability improvement, as well as the feasibility, are also confirmed therein. Section V concludes the paper.

## II. Methodology

### A. Deep learning Koopman operator

A general discrete-time nonlinear system can be defined as:

$$\boldsymbol{x}_{k+1} = \boldsymbol{f}\left(\boldsymbol{x}_k, \boldsymbol{u}_k\right) \tag{1}$$

where $\boldsymbol{x}_k \in \mathbf{R}^n$ is the state vector at time $k$, $\boldsymbol{u}_k \in \mathbf{R}^m$ is the control input vector, and $\boldsymbol{f} : \mathbf{R}^{n+m} \rightarrow \mathbf{R}^n$ is a smooth vector field of $\mathbf{C}^\infty$.

Koopman operator, denoted by $\kappa$, is an infinite linear operator defined on the space of embedding functions $\boldsymbol{g}$ as:

$$\kappa\boldsymbol{g}\left(\boldsymbol{x}_k, \boldsymbol{u}_k\right) = \boldsymbol{g}\left(\boldsymbol{x}_{k+1}, \boldsymbol{u}_{k+1}\right) = \boldsymbol{g}\left(\boldsymbol{f}\left(\boldsymbol{x}_k, \boldsymbol{u}_k\right), \boldsymbol{u}_{k+1}\right) \tag{2}$$

where $\boldsymbol{g} : \mathbf{R}^{m+n} \rightarrow \mathbf{R}^d$ is a smooth vector field of $\mathbf{C}^\infty$.

Therefore, $\boldsymbol{g}$ has lifted the state space to the designated embedding space, where the nonlinear dynamics tend to be linear with Koopman operator $\kappa$. The embedding function $\boldsymbol{g}$ can be further separated into two parts in terms of states and control input.

$$\boldsymbol{g}\left(\boldsymbol{x}_k, \boldsymbol{u}_k\right) = \begin{bmatrix} \boldsymbol{g}_{\boldsymbol{x}}\left(\boldsymbol{x}_k\right) \\ \boldsymbol{u}_k \end{bmatrix} \tag{3}$$

where $\boldsymbol{g}_{\boldsymbol{x}} : \mathbf{R}^n \rightarrow \mathbf{R}^{d-m}$.

The infinite-dimensional $\kappa$ is impossible to realize in real applications but can be truncated with a finite-dimensional approximation [16]. Then, without loss of generality, (2) is simplified within finite dimension space as [21]:

$$\begin{bmatrix} \boldsymbol{g}_{\boldsymbol{x}}\left(\boldsymbol{x}_{k+1}\right) \\ \boldsymbol{u}_{k+1} \end{bmatrix} = \begin{bmatrix} \boldsymbol{K}_{xx} & \boldsymbol{K}_{xu} \\ \boldsymbol{K}_{ux} & \boldsymbol{K}_{uu} \end{bmatrix} \begin{bmatrix} \boldsymbol{g}_{\boldsymbol{x}}\left(\boldsymbol{x}_k\right) \\ \boldsymbol{u}_k \end{bmatrix} \tag{4}$$

where $\boldsymbol{K} \in \mathbf{R}^{d\times d}$ is the finite-dimensional approximation of $\kappa$ in the lifted space.

From (4), we denote $\boldsymbol{K}_{xx} = \boldsymbol{A}$, $\boldsymbol{K}_{xu} = \boldsymbol{B}$, and the state evolution in the lifted space can be written as:

$$\boldsymbol{g}_{\boldsymbol{x}}\left(\boldsymbol{x}_{k+1}\right) = \boldsymbol{A}\boldsymbol{g}_{\boldsymbol{x}}\left(\boldsymbol{x}_k\right) + \boldsymbol{B}\boldsymbol{u}_k \tag{5}$$

where $\boldsymbol{A} \in \mathbf{R}^{(d-m)\times(d-m)}$ and $\boldsymbol{B} \in \mathbf{R}^{(d-m)\times m}$ are the finite-dimensional approximation of $\kappa$ in terms of states. Besides, the aforementioned $\boldsymbol{g}$ in (3) enforces (5) be a control affine system in the lifted space, which is advantageous for future controller design work.

To restore the initial state space from the lifted space at time $k$, $\boldsymbol{g}_{\boldsymbol{x}}$ is defined as:

$$\boldsymbol{g}_{\boldsymbol{x}}\left(\boldsymbol{x}_k\right) = \begin{bmatrix} \boldsymbol{x}_k \\ \boldsymbol{\Phi}_k\left(\boldsymbol{x}_k\right) \end{bmatrix} \tag{6}$$

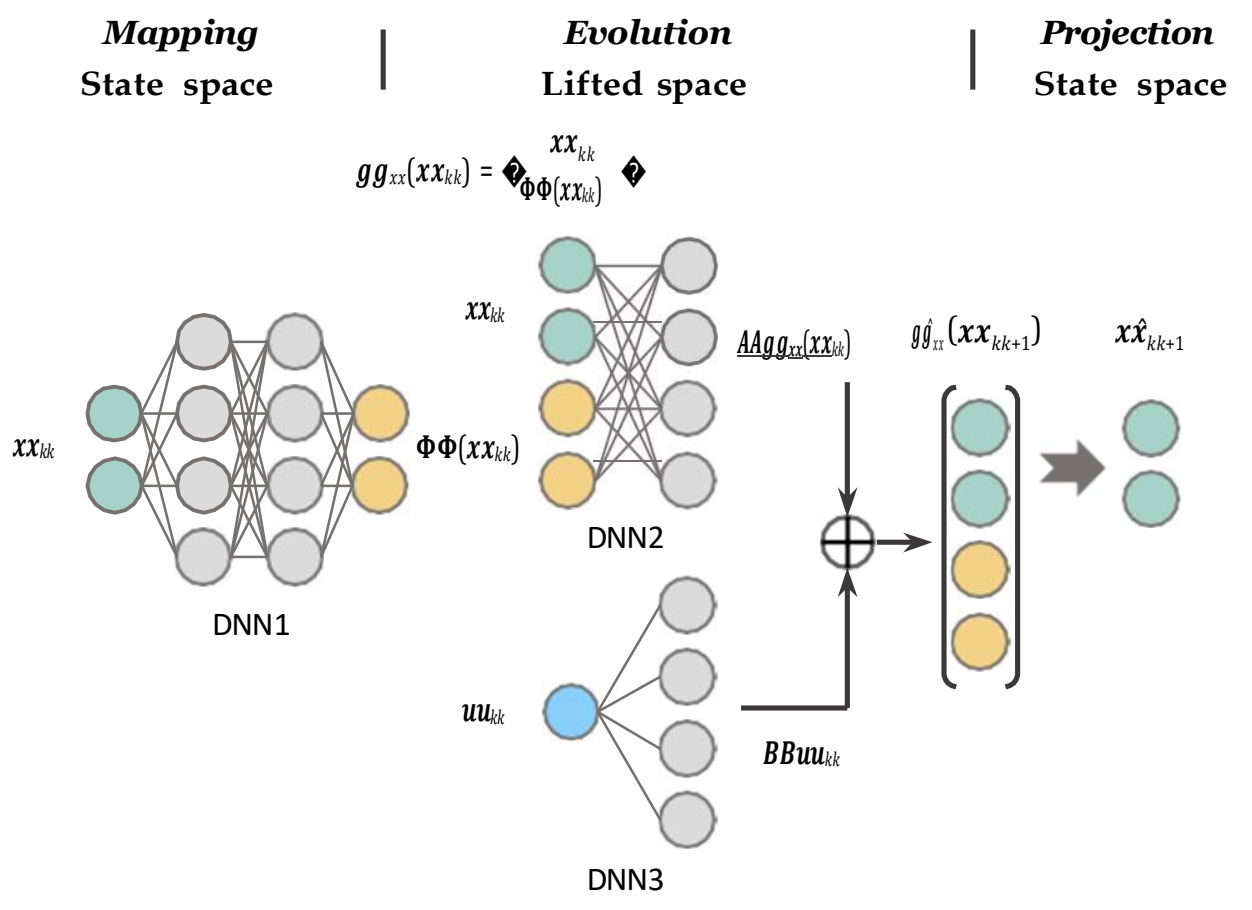

Fig. 1. The framework of deep learning Koopman operator.

where

$$\mathbf{\Phi}_k(\boldsymbol{x}_k) = [\varphi_{1,k}(\boldsymbol{x}_k), \varphi_{2,k}(\boldsymbol{x}_k), \ldots, \varphi_{d-m-n,k}(\boldsymbol{x}_k)]^T \in \mathbf{R}^{d-m-n}$$

is the basis of the lifted space. As such, the state space is included in the lifted space and can be reconstructed without loss of accuracy by the $\boldsymbol{g_x}$ projection onto it.

Deep neural network (DNN) is cast as a collection of nonlinear basis functions. Therefore, the basis $\mathbf{\Phi}$ can be directly learned by training DNN from data pairs to enforce a linear dynamic in the lifted space. The framework of deep learning Koopman operator is illustrated in Fig. 1. It is also worth noting that we employ one hidden layer (without activation and bias) as the linear transformation to feature $\boldsymbol{A}$ and $\boldsymbol{B}$ in the lifted space. This enables the continuous derivatives of the loss function in training, thus remaining error backpropagation, which allows a quick convergence of DNN.

As for the loss function of training DNN, based on the collected data pairs, we adopt the weighted $K$-steps prediction errors in [21] as:

$$L = L(\gamma, \boldsymbol{x}_{i+1}, \hat{\boldsymbol{x}}_{i+1}) \quad (i = 1, 2, \ldots K) \tag{7}$$

where $\gamma \in [0, 1]$ is the decay factor, $\hat{\boldsymbol{x}}_{i+1}(i = 1, 2, \ldots K)$ is the $i$-th step forward prediction from the lifted space by deep Koopman operator, denoted as:

$$\hat{\boldsymbol{x}}_{i+1} = \boldsymbol{C}_P[\boldsymbol{A}\boldsymbol{g_x}(\hat{\boldsymbol{x}}_i) + \boldsymbol{B}\boldsymbol{u}_i] \quad (i = 1, 2, \ldots K) \tag{8}$$

where $\boldsymbol{C}_P = [\boldsymbol{I}_n, \mathbf{O}] \in \mathbf{R}^{n\times(d-m)}$ is the projection matrix from the lifted space to the state space. As we mentioned before, due to the original state space is the sub-space of the lifted space, it can be reshaped by the projection in (8) without losing accuracy, instead of approximation in [20], [29].

### B. Control barrier function

A set $\mathcal{C}$ is the safe set if it is defined as the super-level set of continuously differentiable function $h$: $\subset \mathbf{R}^n \to \mathbf{R}$ that yields [22]:

$$\begin{aligned} \mathcal{C} &= \{\boldsymbol{x} \in D \subset \mathbf{R}^n : h(\boldsymbol{x}) \geq 0\} \\ \partial\mathcal{C} &= \{\boldsymbol{x} \in D \subset \mathbf{R}^n : h(\boldsymbol{x}) = 0\} \\ \mathrm{Int}(\mathcal{C}) &= \{\boldsymbol{x} \in D \subset \mathbf{R}^n : h(\boldsymbol{x}) > 0\} \end{aligned} \tag{9}$$

then $h$ is a control barrier function (CBF) if there exists an extended class $\mathcal{K}_\infty$ function $\alpha$ such that for the control affine system (5):

$$\sup_{\boldsymbol{u}\in\boldsymbol{U}} \dot{h}(\boldsymbol{x}) \geq -\alpha(h(\boldsymbol{x})) \tag{10}$$

for all $\boldsymbol{x} \in D$, the extended class $\mathcal{K}_\infty$ function $\alpha : \mathbf{R} \to \mathbf{R}$ is strictly increasing and with $\alpha(0) = 0$. For simplicity, we define $\alpha(h(\boldsymbol{x})) = ah(\boldsymbol{x})$ as a scalar form with $a \in [0, 1]$ in the following discussions.

The existence of CBF $h$ renders the safe set $\mathcal{C}$ is forward invariant, that is, for every initial condition $\boldsymbol{x}(0) = \boldsymbol{x}_0 \in \mathcal{C}$, there exists $\boldsymbol{u} \in \boldsymbol{U}$ such that the system trajectory $\boldsymbol{x}(t) \in \mathcal{C}$. The forward invariance also implies the system is safe with respect to the safe set $\mathcal{C}$. In other words, if we take a safety-critical constraint of the system (5) as the CBF $h$, rather than the soft or hard constraint, the system states will never violate the constraint in case the initial condition satisfies $h$. Furthermore, the existence condition of CBF in (10) can be further stated in the discrete-time system:

$$\exists \boldsymbol{u}_k \in \boldsymbol{U} \text{ s.t. } h(\boldsymbol{x}_{k+1}) - h(\boldsymbol{x}_k) \geq -ah(\boldsymbol{x}_k) \tag{11}$$

where $\boldsymbol{U} \subseteq \mathbf{R}^m$ is the admissible set of control input. Therefore, the forward invariance of the safe set $\mathcal{C}$, as given by the CBF $h$, is then converted to find an appropriate $\boldsymbol{u}$ that holds (11) at each time step.

The command governor at time $k$ is designed to manage the applied control input $\boldsymbol{u}_k$ to be minimal perturbation on the primary control input $\bar{\boldsymbol{u}}_k$, and to enforce the constraints [30], [31]. Then, we have the following optimization problem, including the input constraints and the CBF conditions:

$$\begin{aligned} &\min_{\boldsymbol{u}_k} \|\boldsymbol{u}_k - \bar{\boldsymbol{u}}_k\|_2^2 \\ &\text{s.t.} \begin{cases} h_j(\boldsymbol{x}_{k+1}) - h_j(\boldsymbol{x}_k) \geq -a_j h_j(\boldsymbol{x}_k) \quad (j \in \mathbf{N}^+) \\ \boldsymbol{x}_{k+1} = \boldsymbol{C}_P[\boldsymbol{A}\boldsymbol{g_x}(\boldsymbol{x}_k) + \boldsymbol{B}\boldsymbol{u}_k] \\ \boldsymbol{u}_k \in \boldsymbol{U} \end{cases} \end{aligned} \tag{12}$$

*Remark 1:* In addition to the forward invariance guarantee of the safety set, the form of the CBF is of considerable relevance. In Section II-A, the original nonlinear system has been lifted into the high-dimensional linear space via the Koopman operator. Based on this, (12) becomes a quadratic programming (QP) problem if $h$ is a linear function with respect to $\boldsymbol{x}$, with which both the accuracy and convergence of the solution can be proven.

## III. Deep Koopman Operator-informed Safety Command Governor Design

The framework of the deep Koopman operator-informed safety command governor is illustrated in Fig. 2. $\boldsymbol{I_d}$ is the reference pose of the autonomous vehicle from the higher-level decision-making controller, and $\boldsymbol{I}_c$ is the actual pose measured by onboard sensors. The primary controller can be referred to as human drivers, particularly those who lack experience and are prone to generate inappropriate maneuvers, or these existing controllers while not or just roughly considering safety concerns. It generates the original control input $\bar{\boldsymbol{u}}$ to enforce the vehicle to behave as expected in $\boldsymbol{I_d}$, however, with the risk of safety violations. The command governor is then built based on the linear dynamics informed by the deep Koopman operator in the lifted space to form the applied control input $\boldsymbol{u}$. Moreover, the CBF conditions, in terms of stability, are formulated and implemented as the constraints on $\boldsymbol{u}$, thus rendering the safety property of vehicle motions.

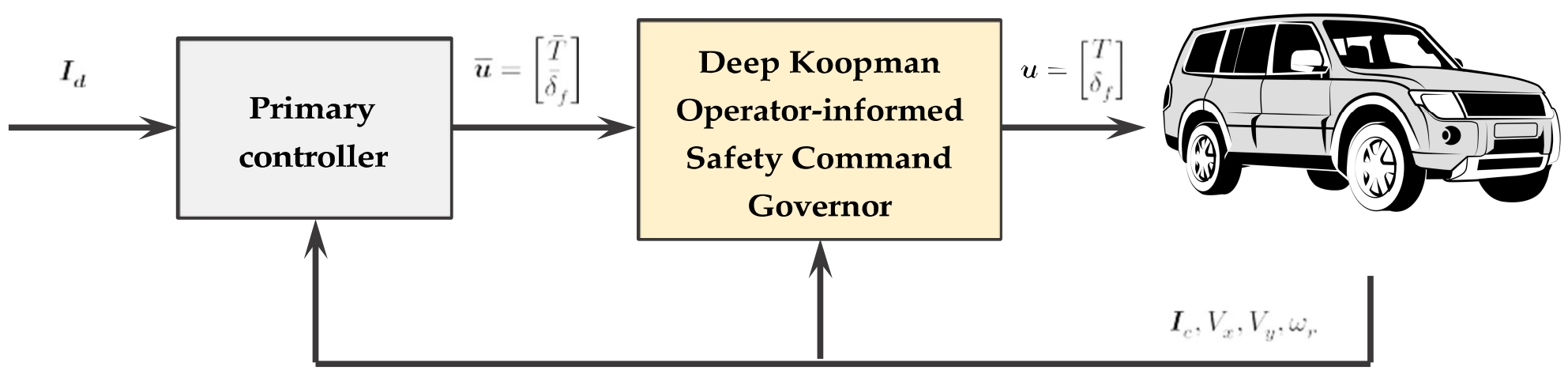

Fig. 2. The framework of deep Koopman operator-informed safety command governor.

### A. Vehicle Planar Dynamics

The planar model is the simplest model in which the vehicle remains parallel to the ground with no roll, no pitch, and no bounce motions. It can feature vehicle dynamic behaviors and predict the dynamic characteristics well [32]. The in-wheel motor driven vehicle (IMDV) is taken as the control plant in this section due to its independent controllability. The four-wheel vehicle planar dynamics model of the IMDV is illustrated in Fig. 3.

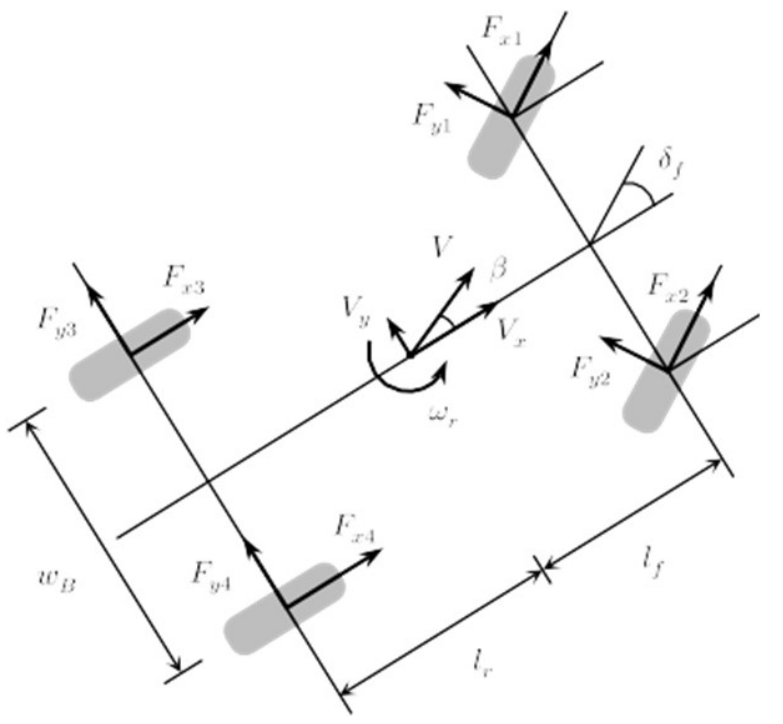

Fig. 3. Four-wheel vehicle planar dynamics model of the IMDV.

The Newton-Euler equations of planar motion for the IMDV are:

$$\begin{cases} \dot{V}_x = \frac{1}{m}\left[(F_{x1}+F_{x2})\cos\delta_f - (F_{y1}+F_{y2})\sin\delta_f\right] \\ \quad + \frac{1}{m}(F_{x3}+F_{x4}) + V_y\omega_r \\ \dot{V}_y = \frac{1}{m}\left[(F_{x1}+F_{x2})\sin\delta_f + (F_{y1}+F_{y2})\cos\delta_f\right] \\ \quad + \frac{1}{m}(F_{y3}+F_{y4}) - V_x\omega_r \\ \dot{\omega}_r = \frac{w_B}{2I_z}\left[(F_{x2}\cos\delta_f - F_{y2}\sin\delta_f) + F_{x4}\right] \\ \quad + \frac{w_B}{2I_z}\left[-(F_{x1}\cos\delta_f - F_{y1}\sin\delta_f) - F_{x3}\right] \\ \quad + (F_{x2}\sin\delta_f + F_{y2}\cos\delta_f + F_{x1}\sin\delta_f + F_{y1}\cos\delta_f)\frac{l_f}{I_z} \\ \quad - (F_{y3}+F_{y4})\frac{l_r}{I_z} \end{cases} \tag{13}$$

where $V_x$, $V_y$, and $\omega_r$ are the longitudinal velocity, lateral velocity and yaw rate of the IMDV, respectively, $F_{xi}(i = 1, 2, 3, 4)$ is the longitudinal tire force, $F_{yi}(i = 1, 2, 3, 4)$ is the lateral tire force, $\delta_f$ is the steering angle of the front wheel, $m$ is the vehicle mass, $I_z$ is the mass moment, $l_f$ is the distance from the front axle to c.g., $l_r$ is the distance from the rear axle to c.g., and $w_B$ is the track width.

The longitudinal tire force can be derived from the equation of rotation motion of the driven wheel as:

$$F_{xi} = F_{xi}(T_i, w_i, F_{zi}, \mu) \quad (i = 1, 2, 3, 4) \tag{14}$$

where $w_i(i = 1, 2, 3, 4)$ is the rotational velocity, $T_i(i = 1, 2, 3, 4)$ is the motor torque or braking torque, $T_{fi}(i = 1, 2, 3, 4)$ is the rolling resistance moment, $F_{zi}(i = 1, 2, 3, 4)$ is the tire vertical load, and $\mu$ is the road friction coefficient, and the lateral tire force is from the following function:

$$F_{yi} = F_{yi}(\alpha_i, F_{zi}, \mu) \quad (i = 1, 2, 3, 4) \tag{15}$$

where $\alpha_i(i = 1, 2, 3, 4)$ is the tire sideslip angle related to velocities, yaw dynamics, steering angle, and vehicle specifications [33].

*Remark 2:* The commonly used linear approximation of (15) [33] is applied for most normal practical driving conditions. However, for extreme scenarios, e.g., icy roads, due to the limited road frictional force, the lateral tire force tends to saturation with increasing tire slip angle; therefore, the linear model is insufficient to depict such nonlinear behaviors. Researchers develop nonlinear empirical models, such as magic formula [34], LuGre [35], etc., to portray the lateral tire force holistically. Nevertheless, it is still challenging to predict the nonlinear behaviors of the tire precisely in all conditions, as the response of lateral tire force is a highly nonlinear component.

The driving torque and the steering angle are the control inputs, and the longitudinal velocity, lateral velocity, and yaw rate are states of the vehicle planar dynamics model:

$$\begin{aligned} \boldsymbol{u} &= [T, \delta_f]^T \\ \boldsymbol{x} &= [V_x, V_y, \omega_r]^T \end{aligned} \tag{16}$$

where the driving torque $T = T_1 + T_2 + T_3 + T_4$ is the summation of motor torque or braking torque of each wheel.

Based on (14)~(16), (13) in the discrete-time manner is denoted as:

$$\begin{bmatrix} V_{x,k+1} \\ V_{y,k+1} \\ \omega_{r,k+1} \end{bmatrix} = \boldsymbol{f}\left( \begin{bmatrix} V_{x,k} \\ V_{y,k} \\ \omega_{r,k} \end{bmatrix}, \begin{bmatrix} T_k \\ \delta_{f,k} \end{bmatrix} \right) \tag{17}$$

where $\boldsymbol{f}$ maps the nonlinear evolution dynamics of the IMDV in a planar motion consistent with that of (1).

### B. Deep learning Koopman operator mapping network

The deep neural network is employed to feature the mapping from the original state space to the lifted space (DNN1) and the linear evolution dynamics in the lifted space (DNN2 for the system matrix, DNN3 for the input matrix). The DNN is comprised of an input layer, hidden layers, and an output layer [36]. Each layer has several neurons or nodes to manage data as the nervous system of living beings [37]. As a supervised neural network, the number of neurons and hidden layers have an impact on $\boldsymbol{\Phi}$ and $\boldsymbol{g_x}$, which subsequently affects the linearity of the lifted space and, in turn, the accuracy of the deep Koopman model. All these factors have the potential to shape the performance of the safety command governor. The details are discussed in the following section.

*1) Data acquisition and processing:* The dataset should consist of diverse road segments and various driving maneuvers to facilitate the network's coverage of the feature space and enhance its generalization performance. On the other hand, the sampling frequency should align with that of sensors and vehicular controllers. Increasing the sampling frequency has the potential to escalate the computing burden and may violate the sampling theorem. On the contrary, a decrease in sampling frequency causes a response delay for the command governor. As such, the collected data were recorded with a sampling interval of 0.05 s, including wheel torque commands, steering wheel angle, longitudinal velocity, lateral velocity (at the center of gravity), and yaw rate.

The data normalization process is conducted in "Min-Max" so that features are on a similar scale before training and end up with smaller standard deviations.

*2) Network training:* The neural network needs to be trained by adjusting the parameters, including weights and bias, to minimize a loss function [10], [38]. As stated in (7), the loss function to characterize the accuracy of vehicle dynamics model in the lifted space is:

$$L = L_1 + L_2 \tag{18}$$

TABLE I
THE DETAILED CONFIGURATIONS OF 3 DNNS

| | DNN1 | DNN2 | DNN3 |
|---|---|---|---|
| Hidden layer nodes | [3 128 128 128 12] | 15 | 2 |
| Batch size / Learning ratio | 20 / 1e-3 | | |
| Sequence length | 20 | | |
| Activation function | ReLu | – | |

In (18) $L_1$ is the prediction loss:

$$L_1(\boldsymbol{\theta}_{\boldsymbol{\Phi}}; \boldsymbol{\theta}_{\boldsymbol{A}}; \boldsymbol{\theta}_{\boldsymbol{B}}) = \sum_{i=1}^{K} \gamma^{i-1} \left\| \begin{bmatrix} V_{x,i+1} \\ V_{y,i+1} \\ \omega_{r,i+1} \end{bmatrix} - \begin{bmatrix} \hat{V}_{x,i+1} \\ \hat{V}_{y,i+1} \\ \hat{\omega}_{r,i+1} \end{bmatrix} \right\|_2^2 \tag{19}$$

where $K$ is the sequence length, $\boldsymbol{\theta}_{\Phi}$, $\boldsymbol{\theta}_{\boldsymbol{A}}$, and $\boldsymbol{\theta}_{\boldsymbol{B}}$ are the learning parameters of DNNs, respectively. Notably, a smaller $\gamma$ places greater emphasis on earlier time steps, while a larger $\gamma$ incorporates all time steps within the prediction horizon. In this case, $\gamma$ is equal to 1.

Further, $L_2$ is the reconstruction loss:

$$L_2(\boldsymbol{\theta}_{\boldsymbol{\Phi}^{-1}}) = \sum_{i=1}^{K} \gamma^{i-1} \left\| \begin{bmatrix} V_{x,i+1} \\ V_{y,i+1} \\ \omega_{r,i+1} \end{bmatrix} - \boldsymbol{\Phi}^{-1}\left( \boldsymbol{\Phi}\left( \begin{bmatrix} V_{x,i+1} \\ V_{y,i+1} \\ \omega_{r,i+1} \end{bmatrix} \right) \right) \right\|_2^2 \tag{20}$$

where $\boldsymbol{\Phi}^{-1}$ is the decoder net of $\boldsymbol{\Phi}$ to enforce the encoder net to capture the main facets of the input data such that the decoder can reconstruct the input based on the encoded features.

The training dataset is from the experimental results in Section IV. The steepest descent is chosen for training based on the gradient of the loss function with respect to the network parameters. The detailed configurations of 3 DNNs are summarized in Table I.

*Remark 3:* The offline training strategy, although it offers a certain degree of generalization, relies heavily on the completeness of the dataset to assure network accuracy. Therefore, it is recommended to include extensive scenarios during the data collection. Alternatively, consider online adjustments to the system matrix $\boldsymbol{A}$ and input matrix $\boldsymbol{B}$ in the lifted space to adapt to vehicle condition changes.

### C. Control barrier functions for vehicle lateral stability

The phase portrait method is regarded as the most intuitive and efficient way to analyze lateral stability, where the stability region is depicted as the combinations of different states, for example, lateral velocity and yaw rate. Based on the research work in [39], we simplify the lateral stability region as a quadrilateral shape in Fig. 4.

The stable region is expected to be forward invariance in terms of safety property, that is, if $(\omega_{r,0}, V_{y,0}) \in \boldsymbol{\Omega}$, then $(\omega_{r,k}, V_{y,k}) \in \boldsymbol{\Omega}$ for $\forall k$. Therefore, the safe set $\boldsymbol{\Omega}$ can be represented as the intersections of four half space as:

$$\boldsymbol{\Omega} = \bigcap_{j=1}^{4} \boldsymbol{\Omega}_j \tag{21}$$

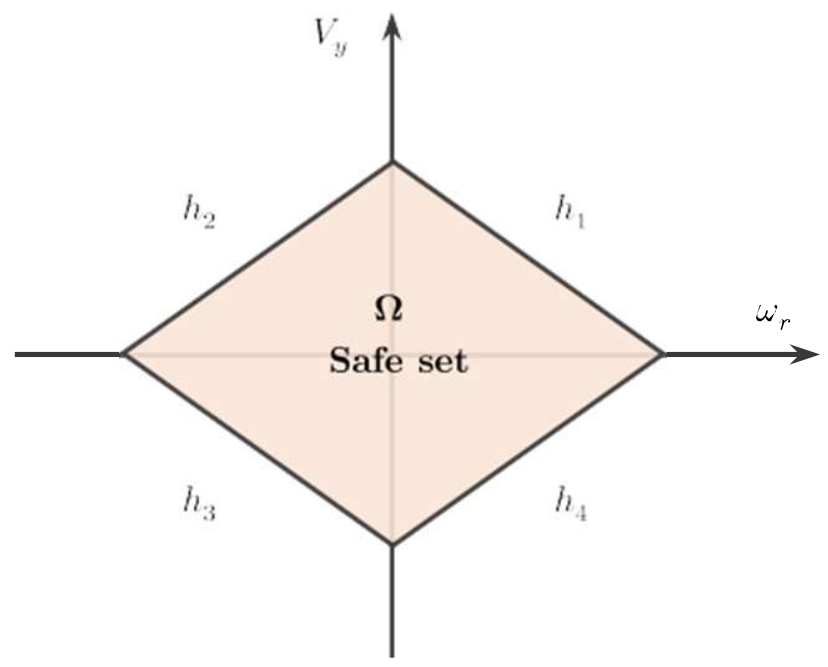

Fig. 4. The lateral stability region (safe set) of the IMDV.

where the half space $\boldsymbol{\Omega}_j (j = 1, 2, 3, 4)$ is:

$$\begin{aligned}
\boldsymbol{\Omega}_1 &: -c_{1,y} V_{y,k+1} - c_{1,r}\omega_{r,k+1} \le b_1 \\
\boldsymbol{\Omega}_2 &: -c_{2,y} V_{y,k+1} - c_{2,r}\omega_{r,k+1} \le b_2 \\
\boldsymbol{\Omega}_3 &: -c_{3,y} V_{y,k+1} - c_{3,r}\omega_{r,k+1} \ge b_3 \\
\boldsymbol{\Omega}_4 &: -c_{4,y} V_{y,k+1} - c_{4,r}\omega_{r,k+1} \ge b_4
\end{aligned} \tag{22}$$

where $c_{j,y}$, $c_{j,r}$, and $b_j (j = 1, 2, 3, 4)$ are the coefficients of the half plane. However, it is another challenging work to obtain values of these parameters that regulate the performance of safety command governor. As outlined in [39], $c_{j,y}$, $c_{j,r}$, and $b_j (j = 1, 2, 3, 4)$ are ought to change with respect to the longitudinal velocity, but we set the stationary safe set for brevity, i.e., $c_{j,y} = 1.3$, $c_{j,r} = 1$, and $b_j = 0.55$ $(j = 1, 2, 3, 4)$.

From (9) and (22), the boundaries of $\boldsymbol{\Omega}$ are denoted as:

$$\begin{cases}
h_1(\boldsymbol{x}) = c_{1,y} V_y + c_{1,r}\omega_r + b_1 \\
h_2(\boldsymbol{x}) = c_{2,y} V_y + c_{2,r}\omega_r + b_2 \\
h_3(\boldsymbol{x}) = -c_{3,y} V_y - c_{3,r}\omega_r - b_3 \\
h_4(\boldsymbol{x}) = -c_{4,y} V_y - c_{4,r}\omega_r - b_4
\end{cases} \tag{23}$$

Further, to maintain the forward invariance of $\boldsymbol{\Omega}$ and enforce $h_j (j = 1, 2, 3, 4)$ to be the CBF, the existence condition at time $k$ is defined as:

$$\begin{aligned}
h_j(\boldsymbol{x}_{k+1}) - h_j(\boldsymbol{x}_k) &\ge -\alpha_j h_j(\boldsymbol{x}_k) \quad (j = 1, 2, 3, 4) \\
h_j(\boldsymbol{x}_{k+1}) &\ge (1 - \alpha_j)\, h_j(\boldsymbol{x}_k) \quad (j = 1, 2, 3, 4)
\end{aligned} \tag{24}$$

We substitute (23) into (24), and the CBF conditions are:

$$\begin{aligned}
& c_{j,y} V_{y,k+1} + c_{j,r}\omega_{r,k+1} + b_j \ge \\
& (1 - \alpha_j)(c_{j,y} V_{y,k} + c_{j,r}\omega_{r,k} + b_j) \quad (j = 1, 2)
\end{aligned} \tag{25a}$$

$$\begin{aligned}
& -c_{j,y} V_{y,k+1} - c_{j,r}\omega_{r,k+1} - b_j \ge \\
& (1 - \alpha_j)(-c_{j,y} V_{y,k} - c_{j,r}\omega_{r,k} - b_j) \quad (j = 3, 4)
\end{aligned} \tag{25b}$$

Considering the CBF conditions and actuation constraints, the command governor generates the applied control input by modifying the original control input of the primary controller with minimal effort. Since the modification on the steering angle often combines with the external path planning algorithm, which is beyond the topic of this paper, we only consider the adjustment on the original driving torque $\bar{T}_k$. This indicates that the original steering angle $\bar{\delta}_k$ will not change, i.e., the applied steering angle $\delta_k$ is equal to $\bar{\delta}_k$ at time $k$. As such, the applied driving torque $T_k$ is the decisive variable of the constrained optimization problem in (26).

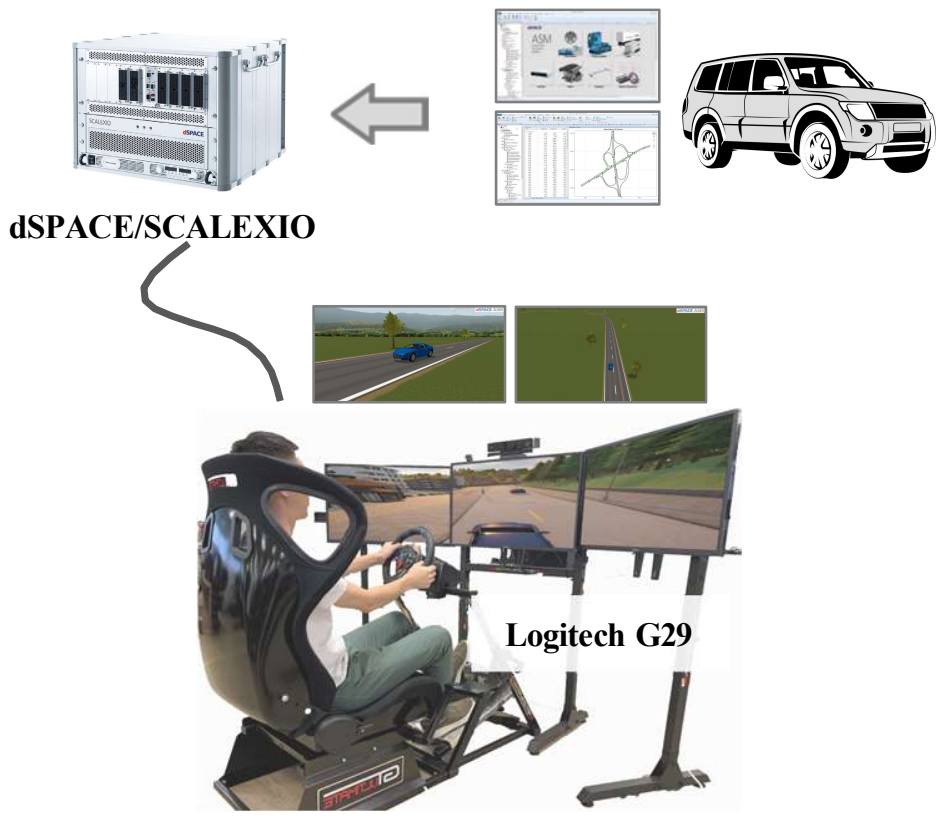

Fig. 5. The hardware-in-the-loop (HiL) platform of dSPACE/SCALEXIO and Logitech G29.

$$\begin{aligned}
& \min_{T_k} \left\| T_k - \bar{T}_k \right\|_2^2 \\
& \text{s.t.} \\
& \boldsymbol{H}_{k+1}\begin{bmatrix} V_{x,k+1} \\ V_{y,k+1} \\ \omega_{r,k+1} \end{bmatrix} - \boldsymbol{H}_k \begin{bmatrix} V_{x,k} \\ V_{y,k} \\ \omega_{r,k} \end{bmatrix} + \boldsymbol{b}_k \ge \mathbf{0}^4 \\
& \begin{bmatrix} V_{x,k+1} \\ V_{y,k+1} \\ \omega_{r,k+1} \end{bmatrix} = \begin{bmatrix} \boldsymbol{I}_3 & \mathbf{0} \end{bmatrix} \left( \boldsymbol{A} g_{\boldsymbol{x}} \left( \begin{bmatrix} V_{x,k} \\ V_{y,k} \\ \omega_{r,k} \end{bmatrix} \right) + \boldsymbol{B} \begin{bmatrix} T_k \\ \delta_k \end{bmatrix} \right) \\
& T_{\min} \le T_k \le T_{\max} \\
& \delta_k = \bar{\delta}_k
\end{aligned} \tag{26}$$

where

$$\boldsymbol{H}_{k+1} = \begin{bmatrix} 0 & c_{1,y} & c_{1,r} \\ 0 & c_{2,y} & c_{2,r} \\ 0 & -c_{3,y} & -c_{3,r} \\ 0 & -c_{4,y} & -c_{4,r} \end{bmatrix}, \boldsymbol{b}_k = \begin{bmatrix} \alpha_1 b_1 \\ \alpha_2 b_2 \\ -\alpha_3 b_3 \\ -\alpha_4 b_4 \end{bmatrix},$$

$$\boldsymbol{H}_k = \begin{bmatrix} 0 & (1-\alpha_1)\, c_{1,y} & (1-\alpha_1)\, c_{1,r} \\ 0 & (1-\alpha_2)\, c_{2,y} & (1-\alpha_2)\, c_{2,r} \\ 0 & -(1-\alpha_3)\, c_{3,y} & -(1-\alpha_3)\, c_{3,r} \\ 0 & -(1-\alpha_4)\, c_{4,y} & -(1-\alpha_4)\, c_{4,r}, \end{bmatrix}$$

$T_{\min}$ and $T_{\max}$ are the lower and upper bound of the driving torque due to motor performance, respectively.

*Remark 4:* As mentioned in Remark 1, $h_j (j = 1, 2, 3, 4)$ has a linear form with respect to $\boldsymbol{x}$, and therefore (26) is a QP problem, which can be solved by means of well-known algorithms [40], [41] and reduce computational efforts. In this case, we adopt YALMIP-IPOPT solver [42] to find the solution.

Besides, we define "one-step prediction," predicting longitudinal velocity, lateral velocity, and yaw rate one step further and comparing them with actual states from dSPACE/SCALEXIO to evaluate the model fidelity.

TABLE II
ROAD SEGMENTS AND FRICTIONAL SURFACE OF COLLECTION DATASET

| **Road segments** | Straight line | Curved roads | | |
|---|---|---|---|---|
| | | $R$=50 m | $R$=75 m | $R$=100 m |
| Length (m) | 1350 | 500 | 200 | 861 |
| **Frictional surface** | $\mu$=0.85 | $\mu$=0.7 | $\mu$=0.5 | $\mu$=0.2 |
| Length (m) | 1861 | 350 | 300 | 400 |

TABLE III
RANGES OF COLLECTION DATASET

| Signals | $T$ (Nm) | $\delta_f$ (rad) | $V_x$ (m/s) | $V_y$ (m/s) | $\omega_r$ (rad/s) |
|---|---|---|---|---|---|
| Max | 847 | 0.11 | 19.91 | 0.29 | 0.37 |
| Min | -2919 | 0.09 | 8.41 | -0.50 | -0.38 |

## IV. VALIDATION AND DISCUSSION

Experimentations are implemented in the hardware-in-the-loop (HiL) platform (Fig. 5) of dSPACE/SCALEXIO and Logitech G29 to collect the training and testing datasets and evaluate the feasibility and effectiveness. To comprehensively explore the feature space both in linear and nonlinear regions, we construct a road layout consisting of straight sections and various curved segments with different road adhesion as listed in Table II, encompassing driving maneuvers such as acceleration, braking, and steering. The timestep is 0.05 s, and the ranges of collection signals are provided in Table III. The outputs of dSPACE/SCALEXIO are the benchmarks for comparison to show the model accuracy of the deep Koopman operator-informed model. Four methods are conducted herein:

1) **7DoF-MF**: a nonlinear physical model to depict the vehicle planar dynamics, proposed in [43]. The magic formula is adopted to obtain the lateral tire forces as the function of tire slip angle and road coefficient. The main parameters of body geometry and tire features are exported from dSPACE/SCALEXIO.
2) **MLP**: a fully connected class of feedforward artificial neural network to predict vehicle planar motions [44]. The input and output configurations align with (16), and the MLP consists of five layers with nodes of [2 128 128 128 3]. Each node in the hidden layers has a nonlinear activation function – ReLu. The batch size is 50, with a learning rate of 0.01 during training.
3) **LSTM**: a recurrent neural network that can process data sequences, such as vehicle accelerations and velocities [44]. The LSTM has feedback connections, of which the weights and biases change once per episode of training, and the activation patterns change once per step. Therefore, the LSTM network is well-suited to cope with time series data in mechatronics systems. Here, the LSTM has the same inputs and outputs as MLP, which exhibits input and output sequence lengths of 5 and 1, respectively. In addition, the LSTM has 3 hidden layers, each comprising 128 cells. During the training phase, a batch size of 20 is employed.
4) **D-Koopman**: DNNs are built to learn the embedding functions and the Koopman operator, as introduced in Section III. The vehicle planar motions are first predicted in the lifted high-dimensional space along the linear evolution dynamics and later projected onto the original space.

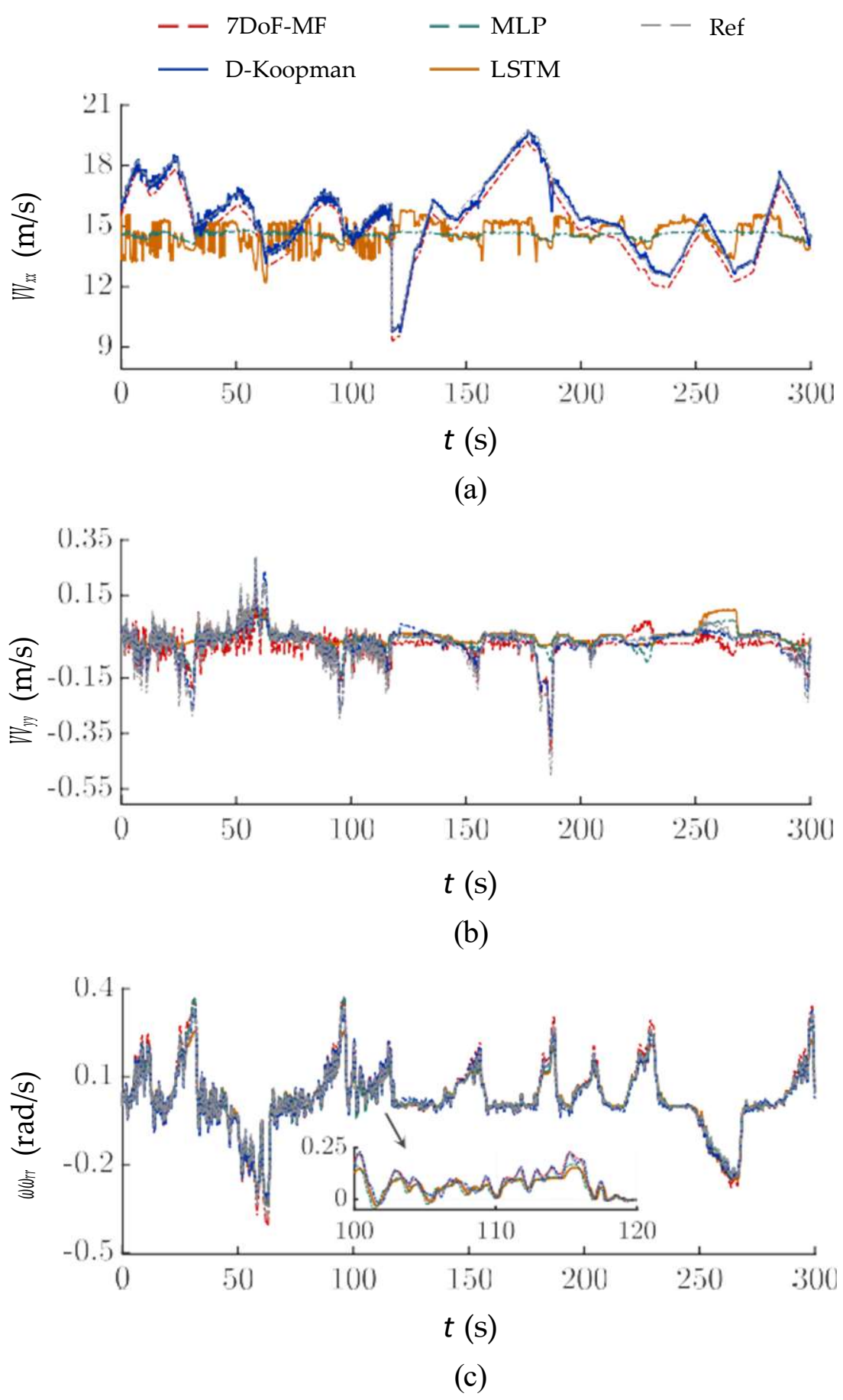

Fig. 6. One-step prediction results an initial longitudinal velocity 50 km/h. (a) Longitudinal velocity. (b) Lateral velocity. (c) Yaw rate.

### A. Modelling accuracy

The evaluation results are illustrated in Fig. 6. From the analysis of the longitudinal velocity prediction results in Fig. 6(a), the 7DoF-MF is able to maintain a similar trend with the actual value but displays a noticeable bias, as evident from the root mean square error (RMSE) value of 0.545 m/s. The MLP and LSTM exhibit ”average” performance on the testing dataset, and the RMSE values are 2.082 m/s and 2.152 m/s, respectively. In contrast, the RMSE value of the D-Koopman model is 0.258 m/s, verifying a higher prediction accuracy.

Moreover, the lateral velocity prediction of the 7DoF-MF has substantial deviation in Fig. 6(b), and the RMSE value is 0.036 m/s. The observed mismatch can be attributed to the physics-based model’s sensitivity to changes in the steering wheel angle, which inhibits its capacity to accurately estimate lateral forces. In contrast, the data-driven models, including the D-Koopman model, exhibit improved performance in pre-dicting lateral velocity, owing to their ability to capture and model complex nonlinear relationships.

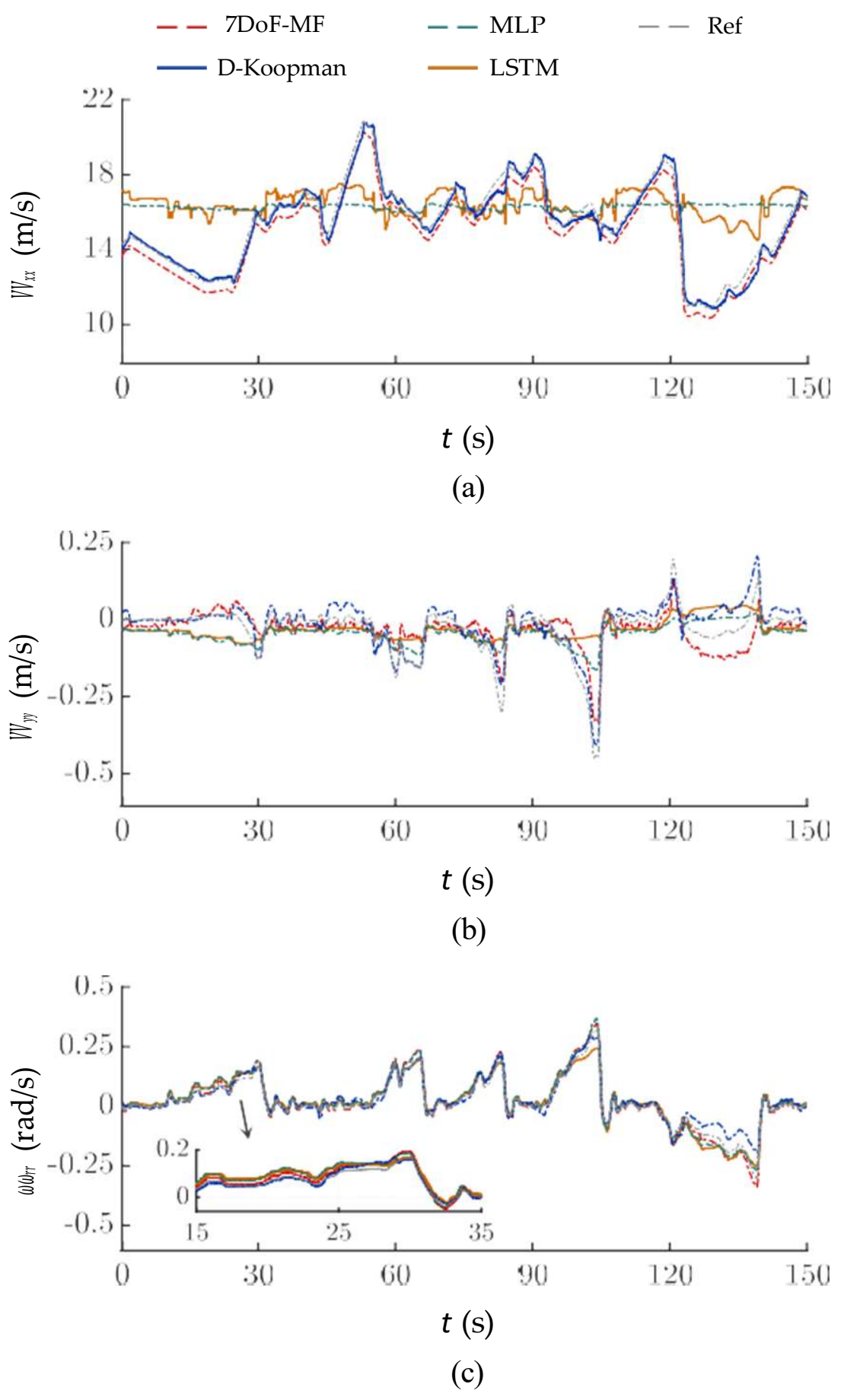

Fig. 7. One-step prediction results with varied mass: $m_0$ + 150 kg. (a) Longitudinal velocity. (b) Lateral velocity. (c) Yaw rate.

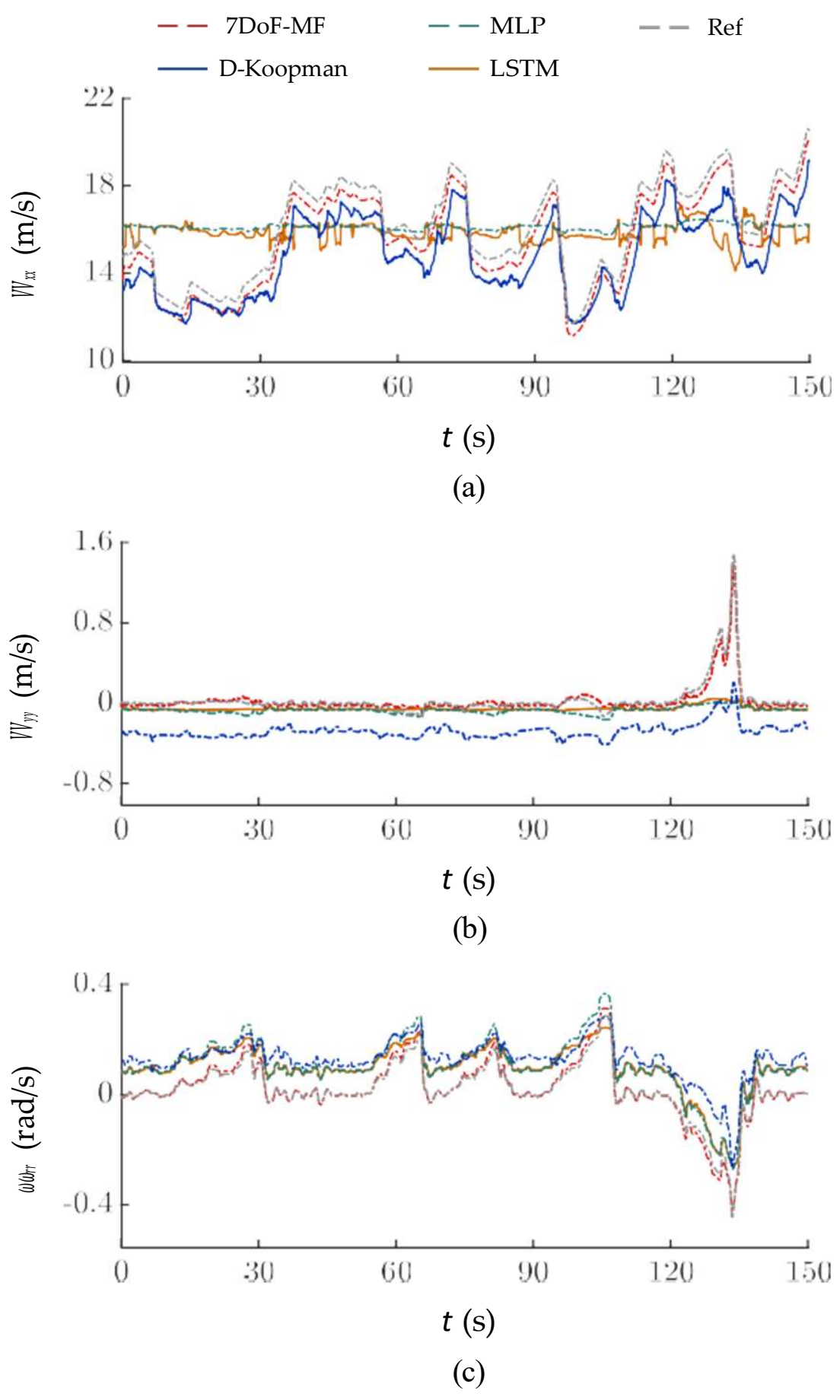

Fig. 8. One-step prediction results with varied mass: $m_0$ - 150 kg. (a) Longitudinal velocity. (b) Lateral velocity. (c) Yaw rate.

As for the yaw rate, the prediction accuracy remains relatively consistent across all four models, especially for the data-driven approaches. This can be attributed to the relatively smaller range of yaw rate values compared to longitudinal and lateral velocities, making it easier for the neural network to establish the relationship between the input and the output features.

The one-step prediction errors for the longitudinal and yaw rate of the 7DoF-MF method are better than those of the deep Koopman model. Moreover, its accuracy of the lateral velocity is sensitive to the mass change, with decreased mass (RMSE = 0.054 m/s) having a greater effect than increased mass (RMSE = 0.038 m/s). In contrast, the D-Koopman model is more robust, with RMSE equal to 0.013 m/s and 0.022 m/s for the lateral velocity, respectively.

In fact, the lateral force induced by tire mechanisms is subject to deformations influenced by external environmental factors, resulting in high nonlinearity. Therefore, the physics-based model (7DoF-MF) cannot precisely predict the lateral force under varying conditions. Besides, the nonlinear physics-based model often consists of a plurality of parameters, and it is challenging to acquire precise value in practical applications. The data-driven methods only collect the input and output data, but the MLP and LSTM models tend to capture an "average" performance due to the loss function. However, the D-Koopman model shows significant superiority in vehicle dynamics modelling, which learns the embedding functions and maps nonlinear dynamics to a linear representation in the lifted space. Moreover, it incorporates more high-order modes of nonlinear components, improving model fidelity than other data-driven methods, and its linear fashion distinguishes it as a preferred choice for elegant model-based technique applications.

To further verify the robustness of the learned deep Koopman model, we change the upsprung mass of the IMVD with ±150 kg around the nominal value of training to conduct the comparison in Fig. 7 and Fig. 8. The initial longitudinal velocity is 50 km/h, and the road adhesion coefficient is 0.85.

The impact of decreased mass on the D-Koopman model's predictions surpasses that of increased mass, emphasizing its sensitivity to changes in vehicle dynamics. Nevertheless, the D-Koopman model continues to outperform the other three models in terms of one-step prediction errors for longitudinal and lateral velocity. Notably, the MLP and LSTM models exhibit marginally higher accuracy in predicting the yaw rate. Additionally, because of the generalization of DNN, the D-Koopman has offset errors in the decreased mass case, but the 7DoF-MF model demonstrates a smaller RMSE value of 0.552 m/s, 0.039 m/s, and 0.013 rad/s for the longitudinal velocity, lateral velocity, and yaw rate, respectively. Nevertheless, the robustness of the D-Koopman model can be strengthened through the collection of a varied range of data or the independent modelling of various uncertainty distributions.

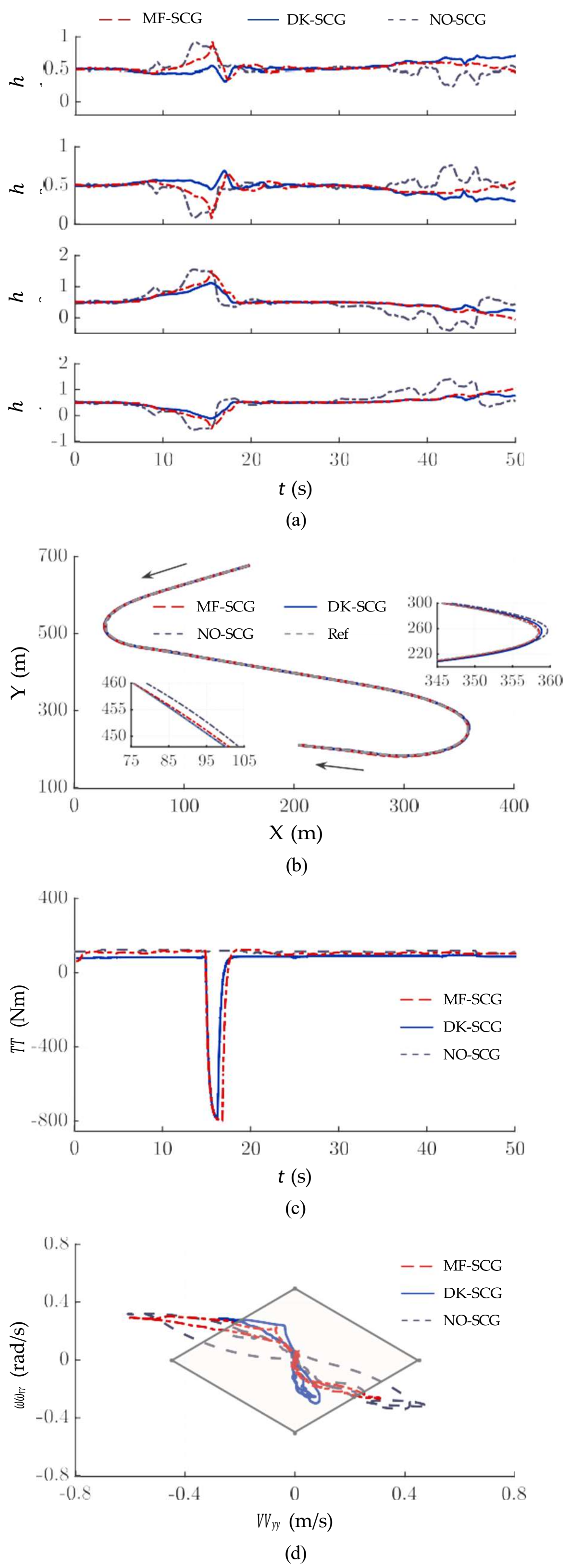

Fig. 9. Validation results on the HiL platform. (a) Control barrier functions. (b) Trajectory. (c) Driving torque. (d) Lateral velocity - Yaw rate phase portrait.

In real-world scenarios, the empirical parameters of physics-based models rely on the specific characteristics of road conditions and vehicle structural parameters. When the environment deviates from the standard calibrated scenarios, these models' parameters lose precision and accuracy. In contrast, the D-Koopman model exhibits greater robustness owing to its capacity for generalization, which stems from the structure and evolution of the deep neural network in the lifted space.

Overall, the D-Koopman model exhibits enhanced accuracy and robustness, surpassing the physics-based model's performance, where nonlinearity becomes more prominent, and offers promising prospects for modeling and predicting vehicle dynamics, providing a valuable alternative to traditional physics-based models. Its ability to capture complex dynamics while maintaining a linear framework presents a significant advantage in terms of computational efficiency and interpretability. By leveraging its generalization capabilities and accommodating an extensive amount of data, the D-Koopman model holds the potential for further refinement and optimization, thus advancing its accuracy and applicability in practical scenarios.

## B. Hardware-in-the-Loop test

We further verify the lateral stability enhancement of deep Koopman operator-informed safety command governor on the HiL platform. The driving torque is evenly distributed to each wheel of the IMDV. Furthermore, we conduct comparisons of three distinct approaches in the following:

1) **NO-SCG**: the applied driving torque is equal to the original driving torque from the primary controller, i.e., no safety-related actions are performed.
2) **MF-SCG**: the similar framework with (26), but the vehicle dynamics evolve in the original state space, as identified by the nonlinear physics-based model, where the magic formula is used to calculate lateral forces. The CBF conditions are also effective and formulate a nonlinear constrained optimization problem to be solved online.
3) **DK-SCG**: the proposed method in (26). Informed by the offline learned deep Koopman operator, the vehicle dynamics has a linear form in the lifted space, and (26) online solves the applied driving torque under the constraints of the CBF conditions.

In Fig. 9(b), it is evident that although all three methods successfully track the desired path, there are notable differences in their performance. Concretely, both the MF-SCG and DK-SCG reduce the driving torque in Fig. 9(c) upon entering the first turn. This torque adjustment enhances handling stability for the subsequent straight road and second turn. Conversely, the NO-SCG traverses the first turn at a higher speed, thus deviating from the desired path, which is later corrected with the assistance of driver maneuvering. Furthermore, during the second turn, the driver must avoid applying large steering wheel angles to maintain vehicle stability, resulting in more noticeable deviations.

On the other hand, both the MF-SCG and DK-SCG modify the original torque command at 14.7 s, and the applied driving torque is less than 0 Nm, which indicates the IMDV is expected to decelerate to mitigate potential risks. Note further that the duration time of torque modification varies due to different model accuracy. The MF-SCG utilizes a physics-based model to predict lateral velocity and yaw rate, whereas the DK-SCG employs a deep Koopman model in a lifted space. The higher accuracy of the deep Koopman model under nonlinear conditions renders it more precise in capturing the vehicle's lateral motions, thus enabling improved intervention and adjustment strategies. As such, compared with the NO-SCG, the phase regions of MF-SCG and DK-SCG become smaller in Fig. 9(d), enhancing the lateral stability of the IMDV. However, the DK-SCG satisfies the CBF conditions ($h_j \geq 0$ ($j = 1, 2, 3, 4$)) in Fig. 9(a), that is, the forward invariance of the safety set. As a result, the DK-SCG promises no violation of the boundaries of safety set in Fig. 9(d), and the IMDV remains in the stable region. This underscores that DK-SCG enforces the safety property of the IMDV through the constraints imposed by CBFs and high model fidelity.

As introduced in Section I, (26) is a QP problem with the embedding functions of the Koopman operator, which helps to decrease its computational effort. The average computational time is 0.1 ms, confirming the applicability and efficiency of the proposed approach.

## V. Conclusion

This paper presents a deep Koopman operator-informed safety command governor for autonomous vehicles. A deep neural network (DNN) is built to extract the kernel functions to approximate the infinite-dimensional Koopman operator. This enables the deep Koopman operator to capture the underlying linear vehicle dynamics in a lifted space, effectively approximating the nonlinearities present in the original state space. The control barrier function is introduced to render the safety property of autonomous vehicles. In terms of lateral stability, the CBF conditions impose constraints on the applied driving torque, and then the QP optimization problem is taken as a safety command governor based on the learned deep Koopman model. Extensive experimentation is conducted on a hardware-in-the-Loop (HiL) platform to collect relevant datasets and validate the proposed method.

The results indicate that the deep Koopman model achieves better accuracy and robustness than the physics-based model – 7DoF-MF, as well as two commonly used data-driven models – MLP and LSTM, across various conditions because of its capacity to feature nonlinear tire forces. Further, the deep Koopman operator–informed safety command governor decreases the original driving torque and maintains the IMDV within the safe set through the constraints imposed by the CBFs constraints and high model fidelity. Consequently, this provides sufficient lateral stability margin to enable the IMDV to follow the desired path. Besides, the linear fashion of the deep Koopman model renders the safety command governor to be a standard QP problem to generate the applied driving torque, thereby minimizing the computational footprint.

Future work should incorporate roll and pitch motions as additional states to develop a more comprehensive data-driven model and discuss the safety command governor to proactively avoid dynamic obstacles.

## References

[1] O. Nelles, *Nonlinear system identification: from classical approaches to neural networks, fuzzy models, and gaussian processes*. Springer Nature, 2020.
[2] J.-P. Noe¨l and G. Kerschen, "Nonlinear system identification in structural dynamics: 10 more years of progress," *Mechanical Systems and Signal Processing*, vol. 83, pp. 2–35, 2017.
[3] A.-T. Nguyen, T.-M. Guerra, C. Sentouh, and H. Zhang, "Unknown input observers for simultaneous estimation of vehicle dynamics and driver torque: Theoretical design and hardware experiments," *IEEE/ASME Transactions on Mechatronics*, vol. 24, no. 6, pp. 2508–2518, 2019.
[4] H. Chen, S. Lou, and C. Lv, "Hybrid physics-data-driven online modelling: Framework, methodology and application to electric vehicles," *Mechanical Systems and Signal Processing*, vol. 185, p. 109791, 2023.
[5] M. Parseh, F. Asplund, L. Svensson, W. Sinz, E. Tomasch, and M. To¨rngren, "A data-driven method towards minimizing collision severity for highly automated vehicles," *IEEE Transactions on Intelligent Vehicles*, vol. 6, no. 4, pp. 723–735, 2021.
[6] L. Liu, Z. Wang, X. Yao, and H. Zhang, "Echo state networks based data-driven adaptive fault tolerant control with its application to electromechanical system," *IEEE/ASME Transactions on Mechatronics*, vol. 23, no. 3, pp. 1372–1382, 2018.
[7] H. Chen and C. Lv, "Rhonn-modeling-based predictive safety assessment and torque vectoring for holistic stabilization of electrified vehicles," *IEEE/ASME Transactions on Mechatronics*, 2022.
[8] S. E. Otto and C. W. Rowley, "Koopman operators for estimation and control of dynamical systems," *Annual Review of Control, Robotics, and Autonomous Systems*, vol. 4, pp. 59–87, 2021.
[9] C. Ren, H. Jiang, C. Li, W. Sun, and S. Ma, "Koopman-operator-based robust data-driven control for wheeled mobile robots," *IEEE/ASME Transactions on Mechatronics*, 2022.
[10] Y. Xiao, X. Zhang, X. Xu, X. Liu, and J. Liu, "Deep neural networks with koopman operators for modeling and control of autonomous vehicles," *IEEE Transactions on Intelligent Vehicles*, 2022.
[11] V. Zinage and E. Bakolas, "Koopman operator based modeling for quadrotor control on se (3)," *IEEE Control Systems Letters*, vol. 6, pp. 752–757, 2021.
[12] G. Mamakoukas, M. L. Castano, X. Tan, and T. D. Murphey, "Derivative-based koopman operators for real-time control of robotic systems," *IEEE Transactions on Robotics*, vol. 37, no. 6, pp. 2173–2192, 2021.
[13] P. J. Schmid, "Dynamic mode decomposition of numerical and experimental data," *Journal of fluid mechanics*, vol. 656, pp. 5–28, 2010.
[14] V. Cibulka, T. Hanisˇ, and M. Hromcˇ´ık, "Data-driven identification of vehicle dynamics using koopman operator," in *2019 22nd International Conference on Process Control (PC19)*. IEEE, 2019, pp. 167–172.
[15] M. O. Williams, M. S. Hemati, S. T. Dawson, I. G. Kevrekidis, and C. W. Rowley, "Extending data-driven koopman analysis to actuated systems," *IFAC-PapersOnLine*, vol. 49, no. 18, pp. 704–709, 2016.
[16] M. O. Williams, I. G. Kevrekidis, and C. W. Rowley, "A data–driven approximation of the koopman operator: Extending dynamic mode decomposition," *Journal of Nonlinear Science*, vol. 25, pp. 1307–1346, 2015.
[17] Y. LeCun, Y. Bengio, and G. Hinton, "Deep learning," *nature*, vol. 521, no. 7553, pp. 436–444, 2015.
[18] S. Kuutti, R. Bowden, Y. Jin, P. Barber, and S. Fallah, "A survey of deep learning applications to autonomous vehicle control," *IEEE Transactions on Intelligent Transportation Systems*, vol. 22, no. 2, pp. 712–733, 2020.
[19] J. Zhang and H. Wang, "Online model predictive control of robot manipulator with structured deep koopman model," *IEEE Robotics and Automation Letters*, 2023.

[20] R. Wang, Y. Han, and U. Vaidya, “Deep koopman data-driven optimal control framework for autonomous racing,” *Early Access*, vol. 5, 2021.

[21] H. Shi and M. Q.-H. Meng, “Deep koopman operator with control for nonlinear systems,” *IEEE Robotics and Automation Letters*, vol. 7, no. 3, pp. 7700–7707, 2022.

[22] A. D. Ames, S. Coogan, M. Egerstedt, G. Notomista, K. Sreenath, and P. Tabuada, “Control barrier functions: Theory and applications,” in *2019 18th European control conference (ECC)*. IEEE, 2019, pp. 3420–3431.

[23] K. Berntorp, R. Quirynen, T. Uno, and S. Di Cairano, “Trajectory tracking for autonomous vehicles on varying road surfaces by friction-adaptive nonlinear model predictive control,” *Vehicle System Dynamics*, vol. 58, no. 5, pp. 705–725, 2020.

[24] J. Theunissen, A. Sorniotti, P. Gruber, S. Fallah, M. Ricco, M. Kvasnica, and M. Dhaens, “Regionless explicit model predictive control of active suspension systems with preview,” *IEEE Transactions on Industrial Electronics*, vol. 67, no. 6, pp. 4877–4888, 2019.

[25] W. Xiao, C. G. Cassandras, and C. Belta, “Decentralized merging control in traffic networks with noisy vehicle dynamics: A joint optimal control and barrier function approach,” in *2019 IEEE Intelligent Transportation Systems Conference (ITSC)*. IEEE, 2019, pp. 3162–3167.

[26] K. Shao, J. Zheng, R. Tang, X. Li, Z. Man, and B. Liang, “Barrier function based adaptive sliding mode control for uncertain systems with input saturation,” *IEEE/ASME Transactions on Mechatronics*, vol. 27, no. 6, pp. 4258–4268, 2022.

[27] J. Liu, H. An, Y. Gao, C. Wang, and L. Wu, “Adaptive control of hypersonic flight vehicles with limited angle-of-attack,” *IEEE/ASME transactions on mechatronics*, vol. 23, no. 2, pp. 883–894, 2018.

[28] A. D. Ames, J. W. Grizzle, and P. Tabuada, “Control barrier function based quadratic programs with application to adaptive cruise control,” in *53rd IEEE Conference on Decision and Control*. IEEE, 2014, pp. 6271–6278.

[29] Y. Han, W. Hao, and U. Vaidya, “Deep learning of koopman representation for control,” in *2020 59th IEEE Conference on Decision and Control (CDC)*. IEEE, 2020, pp. 1890–1895.

[30] I. Kolmanovsky, E. Garone, and S. Di Cairano, “Reference and command governors: A tutorial on their theory and automotive applications,” in *2014 American Control Conference*. IEEE, 2014, pp. 226–241.

[31] E. Garone, S. Di Cairano, and I. Kolmanovsky, “Reference and command governors for systems with constraints: A survey on theory and applications,” *Automatica*, vol. 75, pp. 306–328, 2017.

[32] R. N. Jazar, *Advanced vehicle dynamics*. Springer, 2019.

[33] H. Pacejka, *Tire and vehicle dynamics*. Elsevier, 2005.

[34] M. Bartolozzi, G. Savino, and M. Pierini, “Novel high-fidelity tyre model for motorcycles to be characterised by quasi-static manoeuvres–rationale and numerical validation,” *Vehicle system dynamics*, vol. 60, no. 12, pp. 4290–4316, 2022.

[35] L. Romano, F. Bruzelius, and B. Jacobson, “An extended lugre-brush tyre model for large camber angles and turning speeds,” *Vehicle System Dynamics*, pp. 1–33, 2022.

[36] A. Kurani, P. Doshi, A. Vakharia, and M. Shah, “A comprehensive comparative study of artificial neural network (ann) and support vector machines (svm) on stock forecasting,” *Annals of Data Science*, vol. 10, no. 1, pp. 183–208, 2023.

[37] I. Ullah, K. Liu, T. Yamamoto, R. E. Al Mamlook, and A. Jamal, “A comparative performance of machine learning algorithm to predict electric vehicles energy consumption: A path towards sustainability,” *Energy & Environment*, vol. 33, no. 8, pp. 1583–1612, 2022.

[38] R. Yang, R. Xiong, S. Ma, and X. Lin, “Characterization of external short circuit faults in electric vehicle li-ion battery packs and prediction using artificial neural networks,” *Applied Energy*, vol. 260, p. 114253, 2020.

[39] Y. Huang, S. Z. Yong, and Y. Chen, “Stability control of autonomous ground vehicles using control-dependent barrier functions,” *IEEE Transactions on Intelligent Vehicles*, vol. 6, no. 4, pp. 699–710, 2021.

[40] D. Arnstro¨m and D. Axehill, “A unifying complexity certification framework for active-set methods for convex quadratic programming,” *IEEE Transactions on Automatic Control*, vol. 67, no. 6, pp. 2758–2770, 2021.

[41] P. E. Gill, W. Murray, and M. H. Wright, *Practical optimization*. SIAM, 2019.

[42] J. Lofberg, “Yalmip: A toolbox for modeling and optimization in matlab,” in *2004 IEEE international conference on robotics and automation (IEEE Cat. No. 04CH37508)*. IEEE, 2004, pp. 284–289.

[43] M. Metzler, D. Tavernini, P. Gruber, and A. Sorniotti, “On prediction model fidelity in explicit nonlinear model predictive vehicle stability control,” *IEEE transactions on control systems technology*, vol. 29, no. 5, pp. 1964–1980, 2020.

[44] J. Xu, Q. Luo, K. Xu, X. Xiao, S. Yu, J. Hu, J. Miao, and J. Wang, “An automated learning-based procedure for large-scale vehicle dynamics modeling on baidu apollo platform,” in *2019 IEEE/RSJ International Conference on Intelligent Robots and Systems (IROS)*. IEEE, 2019, pp. 5049–5056.

PLACE PHOTO HERE

**Hao Chen** received the Ph.D. degree in mechanical engineering from School of Vehicle and Mobility, Tsinghua University, China, in 2020.

He is currently a Research Fellow with the School of Mechanical and Aerospace Engineering, Nanyang Technological University, Singapore. His research interests include data-driven modeling, control and optimization of electrified vehicles, safety-critical control theory on mechatronic systems, vehicle dynamics simulation, verification, and control.

PLACE PHOTO HERE

**Xiangkun He** (Member, IEEE) received his PhD degree in 2019 from the School of Vehicle and Mobility, Tsinghua University, Beijing, China. During 2019–2021, he was a Senior Researcher at Noah’s Ark Lab, Huawei Technologies, China.

He is currently a Research Fellow at the School of Mechanical and Aerospace Engineering, Nanyang Technological University, Singapore. His research interests include autonomous vehicles, reinforcement learning, robust machine learning, decision, and control.

PLACE PHOTO HERE

**Shuo Cheng** received the B.S. degree in mechanical engineering from the Harbin Institute of Technology, Harbin, China, in 2016. He received the Ph. D degree in mechanical engineering at the School of Vehicle and Mobility, Tsinghua University, Beijing, China, in 2021. He is currently a JSPS Research Fellow at the Institute of Industrial Science, the University of Tokyo, Tokyo, Japan.

His research interests include intelligent planning and control, automotive chassis system design and dynamics control for automated vehicles.

PLACE PHOTO HERE

**Chen Lv** (Senior Member, IEEE) received the Ph.D. degree from the Department of Automotive Engineering, Tsinghua University, Beijing, China, in January 2016.

He is currently an Assistant Professor with the School of Mechanical and Aerospace Engineering and the Cluster Director in Future Mobility Solutions, NTU. His research focuses on intelligent vehicles, automated driving, and human-machine systems, where he has contributed two books and over 100 articles and obtained 12 granted patents.

Dr. Lv received many awards and honors, selectively including the Highly Commended Paper Award of IMechE U.K. in 2012, the IEEE IV Best Workshop/Special Session Paper Award in 2018, the Automotive Innovation Best Paper Award in 2020, the winner of the Waymo Open Dataset Challenges at CVPR 2021, and the Machines Young Investigator Award in 2022. He serves as an Associate Editor for IEEE Transactions on Intelligent Transportation Systems, IEEE Transactions on Vehicular Technology, and IEEE Transactions on Intelligent Vehicles.